# A Short Survey on Set-Based Aggregation Techniques for Single-Vector WSI Representation in Digital Pathology

S. Hemati[1], Krishna R. Kalari[2], H.R. Tizhoosh[1]

[1] Kimia Lab, Dept. of Artificial Intelligence & Informatics
Mayo Clinic, Rochester, MN, USA

[2] Division of Biomedical Statistics and Informatics, Dept. of Health Sciences Research
Mayo Clinic, Rochester, MN, USA

**Abstract** - Digital pathology is revolutionizing the field of pathology by enabling the digitization, storage, and analysis of tissue samples as whole slide images (WSIs). WSIs are gigapixel files that capture the intricate details of tissue samples, providing a rich source of information for diagnostic and research purposes. However, due to their enormous size, representing these images as one compact vector is essential for many computational pathology tasks, such as search and retrieval, to ensure efficiency and scalability. Most current methods are "patch-oriented," meaning they divide WSIs into smaller patches for processing, which prevents a holistic analysis of the entire slide. Additionally, the necessity for compact representation is driven by the expensive high-performance storage required for WSIs. Not all hospitals have access to such extensive storage solutions, leading to potential disparities in healthcare quality and accessibility. This paper provides an overview of existing set-based approaches to single-vector WSI representation, highlighting the innovations that allow for more efficient and effective use of these complex images in digital pathology, thus addressing both computational challenges and storage limitations.

## Introduction

Whole Slide Images (WSIs) are a critical resource in the field of digital pathology, allowing for the digitization, storage, and detailed analysis of biopsy samples. However, the gigapixel scale of these images poses substantial computational challenges. To address this issue, typically each WSI is split to multiple patches that can be fitted into GPU memory. In this scenario, we end up with a set of patch embeddings per WSI. such representation still is not optimal in terms of memory storage and downstream data analysis tasks. To mitigate these issues, recent research has concentrated on developing methods to aggregate and condense these massive images into compact single-vector representations. These techniques not only facilitate more efficient storage but also enhance the capabilities for classification, retrieval, and analysis of the data. This brief survey delves into the importance of these technological advancements by reviewing the key studies that have significantly contributed to the field.

An alternative approach to condensing or aggregating whole slide image (WSI) embeddings into a single representation is the "median-of-minimums" method, which has been proposed for image retrieval and applied in classification tasks using the k-nearest neighbors (k-NN) approach (Kalra, 2020). In this method, after segmenting the WSI into patches, each patch from the input WSI is compared with all patches from another WSI to assess their similarity (refer to Figure 1). However, to make this process manageable and efficient, it is essential to select a small, representative subset of tiles or patches from each WSI for these comparisons or for aggregation. This is particularly important when the embeddings need to be saved for future use, such as in image retrieval tasks (Tizhoosh, 2024). Several methods have been developed for this purpose, including the "mosaic" method by Yottixel (Kalra, 2020), the "montage" method by SPLICE (Alsaafin, 2024), and the "assembly" method by SDM (Shafique, 2023). Bypassing the patching step and attempting to process all patches for aggregation can lead to significantly slower execution times and increased inefficiency, making the process more expensive in terms of both computation and storage. Thus, careful selection and aggregation of representative patches are crucial to optimizing performance and resource use.

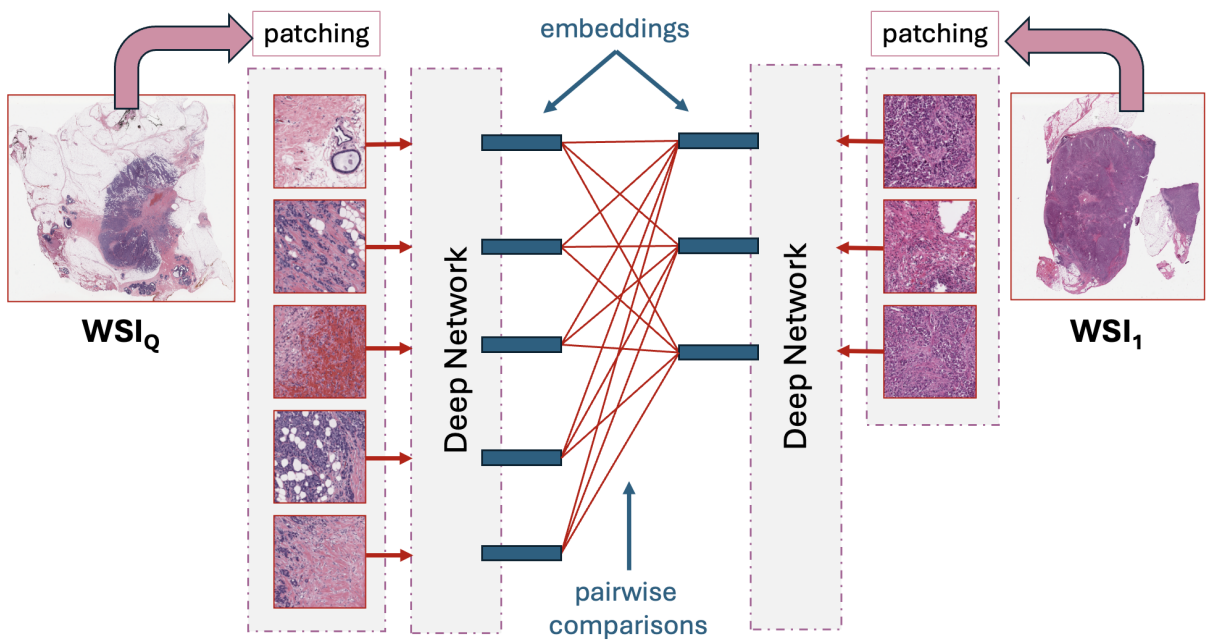

*Figure 1. Yottixel introduced the "median-of-minimums" method to compare two WSIs [Kalra2020].*

Aggregation models, in contrast, combine all embeddings of a WSI into a single vector to streamline image analysis in computational pathology (see Figure 2). This could occur through averaging all embeddings ($SV_{A1}$ in Figure 2 using a model A1), **dependent aggregation** ($SV_{A2}$ in Figure 2 via a model A2 using a pretrained model A1), **independent aggregation** ($SV_B$ in Figure 2 via the model B), or **low power embedding** ($SV_C$ in Figure 2 via the model C trained on the entire WSI in low magnification). Table 1 summarizes the main characteristics of this general approach.

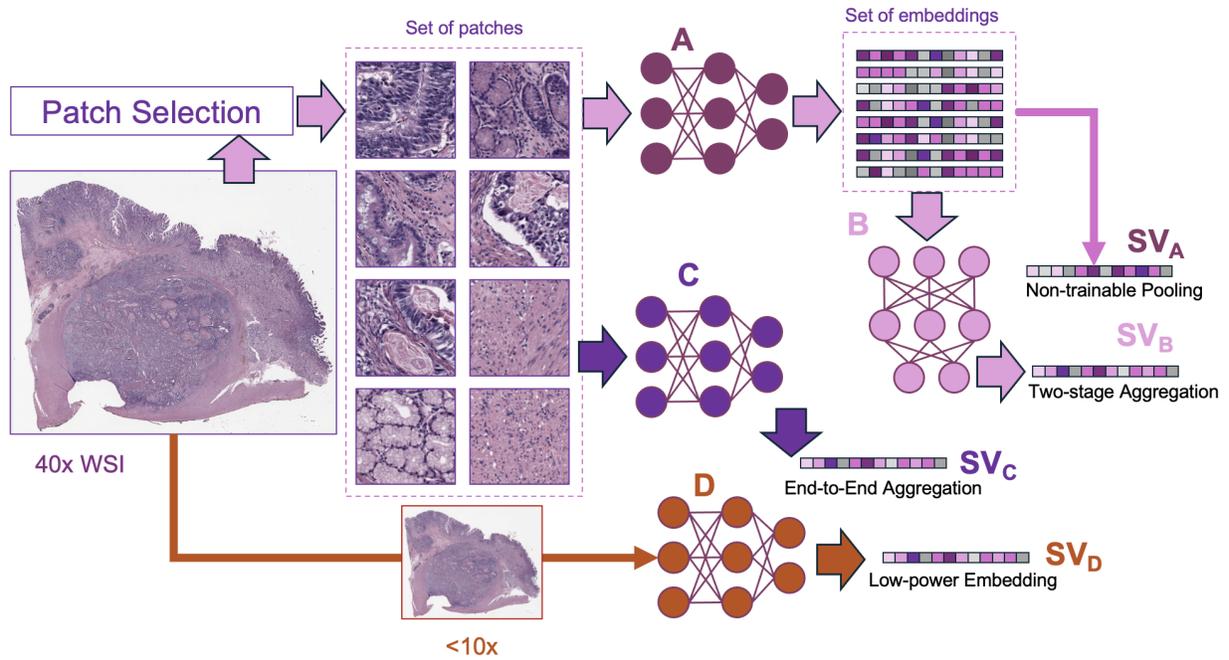

*Figure 2. Different approaches to generating a **single-vector (SV)** to represent a given WSI. Patch selection is a pivotal part of the process to ascertain efficient implementation. Methods like Yottixel's mosaic and SPLICE can deliver a small set of patches in an unsupervised manner. The patches can go through a network A to generate embeddings (feature vectors). A simple pooling function can be used to build a single vector $SV_A$ for the entire WSI. The embeddings may be aggregated by another network B to generate a single vector $SV_B$ as well. A network C may also directly operate on patches to generate a single vector $SV_C$. As another possibility, a low magnification WSI may be fed into a network D to process it as a single image and generate $SV_D$.*

Table 1. The general characteristics of different approaches to single-vector embedding of WSIs

|  | Needs Training | Applicable on Pre-Trained Models | Design Complexity | Expected Accuracy |
|---|---|---|---|---|
| **Pooling** ($SV_A$ in Fig. 1) | No | Yes | Very Low | Low to Medium |
| **Two-stage Aggregation** ($SV_B$ in Fig. 1) | Yes | No | Low | Medium to High |
| **End-to-end Aggregation** ($SV_C$ in Fig. 1) | Yes | No | High | Medium to High |
| **Low Power Embedding** ($SV_D$ in Fig. 1) | Yes | No | Low | Low to Medium |

In the following sections, we briefly review the methods proposed to aggregate embeddings of WSI patches into a single vector. Generally, these methods can be divided into two main categories. Techniques that take into account the spatial information of patches or patch embeddings and the ones that obtain the single-vector embedding without considering spatial information. Considering that the later ones are simpler and usually need less memory and computational load and still achieve similar performance compared to the first category, here we only review the ones that do not consider the spatial information of patches.

**Evolutionary Deep Feature Selection for Compact Representations, (Bidgoli, 2022): Pooling ($SV_A$) + Evolutionary Feature Selection**

In their 2022 study, Bidgoli et al. address the challenge of creating compact WSI representations by proposing an evolutionary deep feature selection approach. The method involves extracting deep feature vectors from WSIs and using a simple average pooling to obtain single WSI embedding per WSI. Having this single vector WSI representation, evolutionary algorithms are used to optimize and condense these vectors into a compact feature vector (CFV). This CFV is significantly smaller—up to 11,000 times—than the original feature set while maintaining high classification accuracy. This reduction in data size facilitates more efficient computational processing and storage, making it easier to handle large volumes of pathology images. The study demonstrates that this compact representation improves classification accuracy by 11% compared to previous benchmarks, highlighting its potential to enhance the efficiency and scalability of digital pathology workflows.

**Set Representation Learning using Memory Networks (Kalra, 2020): Two-stage Aggregation ($SV_B$)**

Kalra et al. developed the Memory-based Exchangeable Model (MEM) concept to learn the permutation invariant representations for the set data. In this method, a MEM is constructed from multiple memory blocks where each memory block is a sequence-to-sequence model containing multiple memory units. The output of each memory block is invariant to the permutations of the input sequence. The memory units in this architecture are in charge of calculation of attention values for each instance. To calculate these attention values, the memory vectors are aggregated using a pooling operation (weighted averaging) to form a permutation-invariant representation. Increasing the number of memory units enables the memory block to capture more complex dependencies between the instances of a set by providing an explicit memory representation for each instance in the sequence. MEM shows promising results on different set classification tasks including point cloud and lung WSIs classification.

**CNN and Deep Sets for WSI representation learning, (Hemati, 2021): End-to-end Aggregation ($SV_C$)**

In the study by (Hemati, 2021) the authors tackle the issue of computational bottlenecks inherent in processing WSIs. They propose a novel neural network architecture that integrates Convolutional Neural Networks (CNNs) and Deep Sets (Zaheer, 2017) as one the first deep Multiple Instance Learning (MIL) techniques to extract a single permutation-invariant vector representation for each WSI. They showed that by incorporating 40 patches per WSI, and training the CNN and Deep Sets units in an end-to-end manner, they can achieve high quality single-vector WSI representations that can outperform earlier multi vector (Kalra, 2020) representations. This approach is particularly significant as it addresses the difficulty of employing deep learning directly on gigapixel images by avoiding the need for patch-based processing. Instead of working with a bag of patches, this method provides a cohesive and compact vector representation that simplifies downstream tasks such as image search and classification. The network is trained in a multi-label setting to encode both primary site and diagnosis, enabling efficient transfer learning and achieving superior performance in retrieval and classification tasks compared to existing methods like Yottixel (Kalra, 2020).

**Attention-based deep multiple instance learning, (Ilse, 2018): End-to-end Aggregation ($SV_C$)**

To tackle limitations of early stage MIL layers including being non-trainable and giving no score to the importance of instances within a bag as in Deep Sets (Zaheer, 2017), Ilse et al. proposed attention based MIL pooling layer. Authors were also motivated given ability to detect more important instances, such a flexible and adaptive MIL pooling could lead superior set level representation and offer better interpretability. Attention based MIL use a weighted average of low-dimensional instance representations where weights must sum to 1 and determined by a simple network. Ilse et al. validated their proposed attention based MIL on two different pathology datasets breast and colon cancer and showed superior performance compared with the rigid max and average pooling.

**CNN with Attention based MIL and Self-Supervised Contrastive Learning, (Fashi, 2022): End-to-end Aggregation ($SV_C$)**

Fashi et al. introduced a self-supervised contrastive learning framework specifically designed for learning end-to-end WSI representation learning. This method leverages the primary site information of WSIs to enhance the robustness of the learned representations. Unlike traditional augmentation-based self-supervised learning approaches, this framework directly focuses on end-to-end WSI representation (rather than patch-based processing) given primary site information as the first training stage. Then, the Network is further fine-tuned using a supervised contrastive loss. By doing so, it improves the generalization capabilities of the model for downstream tasks such as classification and search. In this work, the patch embeddings are aggregated into a single vector using an attention-based MIL layer proposed by (Ilse, 2018). The model was trained and evaluated on over 6,000 WSIs from The Cancer Genome Atlas (TCGA) repository, demonstrating excellent performance across various primary sites and cancer subtypes, particularly excelling in lung cancer classification. This study underscores the potential of self-supervised learning in generating robust and efficient WSI representations.

**Focal Attention for WSI Representation Learning, (Kalra, 2021): Two-stage Aggregation ($SV_B$)**

Another MIL scheme developed for WSI classification/search is focal attention proposed in (Kalra, 2021). In this work, the authors were inspired by the two recent developments in representation learning literature focal loss and attention based MIL and proposed the novel pooling layer called focal attention (FocAtt-MIL). In FocAtt-MIL, the attention-weighted averaging proposed by (Ilse, 2018) is additionally modulated by a trainable focal factor. More precisely, FocAtt-MIL is composed from four main components. Prediction MLP, WSI Context, Attention Module, and Focal Network. The Prediction MLP is a trainable neural network that calculates a prediction for each patch embedding in the set. WSI context is also a neural network (basically Deep Sets from (Zaheer, 2017) that obtains one embedding per WSI in a simple efficient manner that capture a general information from the WSI. The attention module is taken from the to (Ilse, 2018) work which is itself made from two MLPs transformation, and the Attention networks. The attention module takes the patch embedding and WSI context and output the an attention value between 0 to 1 for each instance. Finally, the focal network receive the WSI context and computes a focal factor per dimension which further guide the final prediction towards better WSI representation. FocAtt-MIL has been tested on classification of Lung Adenocarcinoma (LUAD) and Lung Squamous Cell Carcinoma (LUSC) and also Pan-cancer Analysis for a dataset of 7,097 training, and 744 test WSIs covering 24 different anatomic sites, and 30 different primary diagnoses.

**Incorporating intratumoral heterogeneity into weakly-supervised deep learning models via variance pooling (Carmichael, 2022): End-to-end Aggregation ($SV_C$)**

As an improvement to the attention-based pooling, Carmichael et al. proposed to add a variance pooling on top of attention values to capture within-WSI heterogeneity (intratumoral heterogeneity). In this architecture, the variance of attention values is concatenated to the mean values and fed to a MLP to calculate the final WSI embedding. This architecture was evaluated on five cancer types from TCGA and showed adding this variance pooling on attention values improves predictive performance of the model on survival prediction.

**Dual-stream multiple instance learning network for whole slide image classification with self-supervised contrastive learning, (Li, 2021): Two-stage Aggregation ($SV_B$)**

Li et al. proposed Dual-stream multiple instance learning (DSMIL) aggregation technique to capture the dependencies of the instances in a dual-stream architecture with trainable distance measurement. In DSMIL, aggregation is performed using two units namely a masked non-local block and a max-pooling layer (Zaheer, 2017) . Additionally, to mitigate challenges caused by large or unbalanced bags for the training of the models, authors proposed to use self-supervised contrastive learning. In DSMIL, the instance and the bag classifiers are used simultaneously for both streams. The first stream applies max pooling on embeddings obtained through an instance classifier on each instance. The second stream aggregates the above instance embeddings into a bag embedding which is eventually scored by a bag classifier. Finally, an attention-like scoring mechanism is used to detect important instances. Authors tested their approach on Camelyon16 and TCGA lung datasets. In this paper, the two-stage aggregation technique was used. First, the backbone was used in self-supervised manner using SimCLR and then the extracted features were aggregated using the DSMIL.

**ReMix: A general and efficient framework for multiple instance learning based whole slide image classification, (Yang, 2022): Two-stage Aggregation ($SV_B$)**

Authors in ReMix paper target the memory and computational load of recent MIL techniques developed for WSI representation learning task. This is due the large size of WSIs that usually lead to gigantic bags to tens of thousands of elements (patches). Additionally, data augmentation for MIL based WSI representation learning is explored. RemMix involves two stages: reduce and mix. First, the patch prototypes, i.e., patch cluster centroids are detected and used instead of non-prototype patches. Then the bag mixing augmentation step which includes four latent space augmentations is performed. This step diversify discriminative representations in the latent space. The effectiveness of ReMix was validated on UniToPatho and Camelyon16 datasets where authors showed significant improvement over DSMIL by (Li, 2021).

**Deep fisher vector for Sparse and Binary permutation invariant Representations for Real-Time Image Retrieval, (Hemati, 2023): Two-stage Aggregation ($SV_B$)**

In their 2023 study, Hemati et al. propose a framework for learning sparse and binary permutation invariant WSI representations to improve real-time image retrieval and classification performance. The framework leverages deep conditional generative modeling and Fisher Vector Theory to create two types of compact representations: Conditioned Sparse Fisher Vector (C-Deep-SFV) and Conditioned Binary Fisher Vector (C-Deep-BFV). These representations are designed to be memory-efficient and computationally efficient, making them suitable for large-scale medical archives. Given that Fisher Vector is the normalized gradient of generative model loss function with respect to its parameters, to have sparse and binary WSI embeddings the authors introduced new loss functions—gradient sparsity and gradient quantization losses—that encourage gradient space of the generative model to be sparse and binary respectively. This is the first study that obtain sparse or binary permutation WSI embeddings which are efficient for retrieval systems. Validated on datasets from TCGA and the Liver-Kidney-Stomach (LKS) dataset, the proposed method outperforms existing retrieval systems like Yottixel, achieving better retrieval accuracy and speed. Additionally, the framework shows competitive performance in WSI classification tasks, further demonstrating its practical utility in clinical settings.

**Graph convolutional neural network for WSI representation learning, (Adnan, 2020): Two-stage Aggregation ($SV_B$)**

Adnan et.al. proposed to represent each WSI as a fully connected graph where patch embeddings concatenated with a global context vector obtained by a pooling function represent nodes in the graph. Zaheer et.al. In the Deep Sets paper, Zaheer showed that a mapping followed pooling function can be used as a universal set approximator which is used as context vector by Adnan. Having the edges, the Adjacency Matrix is learned using an Adjacency Matrix learning layer. The element in row i and column j of Adjacency Matrix quantifies similarity between nodes i and j. Having the graph representation of each WSI, graph convolution network is used to learn the representation of the WSI as a graph. This representation encodes relations among patch embeddings. To obtain the single vector representation of the WSI, the graph representation is fed to graph pooling layer. Adnan et.al. introduced attention via graph pooling to capture infer patches with higher importance. They tested their proposed technique for classifying lung cancers into Lung Adenocarcinoma (LUAD) & Lung Squamous Cell Carcinoma (LUSC) over 1,026 lung cancer WSIs in 40X magnification and obtained the state-of-the-art accuracy of 88.8%.

## Conclusions

The representation of WSIs as single vectors marks a significant advancement in digital pathology. These methods not only address the computational challenges posed by gigapixel images but also enhance the efficiency and accuracy of downstream tasks such as classification and retrieval. By leveraging deep multi-instance learning, self-supervised learning, and evolutionary algorithms, researchers have been able to obtain WSI embeddings with higher quality. These innovations hold great promise for improving diagnostic accuracy and treatment outcomes in clinical settings. For future work, this is interesting to see quantitative comparison of WSI representation learning techniques in terms of final embedding quality, their applicability for different tasks (classification, retrieval, segmentation, etc.) and computational load in a fair and comprehensive manner.